# Magnetism in heavy-fermion U(Pt,Pd)$_3$ studied by μSR

R.J. Keizer, A. de Visser, A.A. Menovsky, and J.J.M. Franse
*Van der Waals-Zeeman Institute, University of Amsterdam,*
*Valckenierstraat 65, 1018 XE Amsterdam, The Netherlands*

A. Amato
*Paul Scherrer Institute, CH-5232 Villigen, Switzerland*

F.N. Gygax, M. Pinkpank, and A. Schenck
*Institute for Particle Physics, ETH Zürich, PSI, CH-5232 Villigen, Switzerland*

**Abstract**

We report μSR experiments carried out on a series of heavy-electron pseudobinary compounds U(Pt$_{1-x}$Pd$_x$)$_3$ ($x \leq 0.05$). For $x \leq 0.005$ the zero-field muon depolarisation is described by the Kubo-Toyabe function. However the temperature variation of the Kubo-Toyabe relaxation rate $\Delta_{KT}(T)$ does not show any sign of the small-moment antiferromagnetic phase with $T_N \approx 6$ K (signalled by neutron diffraction), in contrast to previous reports. The failure to detect the small ordered moment suggests it has a fluctuating (> 10 MHz) nature, which is consistent with the interpretation of NMR data. For $0.01 \leq x \leq 0.05$ the muon depolarisation in the ordered state is described by two terms of equal amplitude: an exponentially damped spontaneous oscillation and a Lorentzian Kubo-Toyabe function. These terms are associated with antiferromagnetic order with substantial moments. The Knight-shift measured in a magnetic field of 0.6 T on single-crystalline U(Pt$_{0.95}$Pd$_{0.05}$)$_3$ in the paramagnetic state shows two signals for $\mathbf{B} \perp \mathbf{c}$, while only one signal is observed for $\mathbf{B} \parallel \mathbf{c}$. The observation of two signals for $\mathbf{B} \perp \mathbf{c}$, while there is only one muon localisation site (0,0,0) points to the presence of two spatially distinct regions of different magnetic response.

Corresponding author:   Dr. A. de Visser
                        Van der Waals-Zeeman Institute, University of Amsterdam
                        Valckenierstraat 65, 1018 XE Amsterdam, The Netherlands
                        Phone:  31-20-5255732
                        Fax:    31-20-5255788
                        E-mail: devisser@wins.uva.nl

# 1. Introduction

The heavy-fermion material UPt$_3$ continues to attract a great deal of attention, because of its unconventional magnetic and superconducting properties. The low-temperature normal-state of UPt$_3$ [1, 2] presents an exemplary strongly renormalised Fermi-liquid, with a quasiparticle mass of the order of 200 times the free electron mass, as evidenced by the large coefficient of the linear term in the specific heat, $\gamma = 0.42$ J/molK$^2$, and the equally enhanced Pauli susceptibility, $\chi_0(T \rightarrow 0)$. The magnetic properties of this hexagonal material are quite intriguing. The magnetic susceptibility $\chi(T)$ has a broad maximum at $T_{max} \approx 18$ K for a field in the hexagonal plane ($B \rightarrow 0$), which is attributed to the stabilisation of antiferromagnetic interactions below $T_{max}$. For $T < T_{max}$ the magnetisation $\sigma(B)$ exhibits a magnetic transition at a field $B^* = 20$ T (**B** $\perp$ **c**). This continuous phase transition has been termed pseudo-metamagnetic and is interpreted as a suppression of the antiferromagnetic interactions. The most striking magnetic property of UPt$_3$ is undoubtedly the small-moment antiferromagnetic order (SMAF) which develops below the Néel temperature $T_N \approx 6$ K [3]. The ordered moment $m = 0.02$ $\mu_B$/U-atom is unusually small and is directed along the a$^*$ axis in the hexagonal plane. The magnetic unit cell consists of a doubling of the nuclear unit cell along the a$^*$ axis. This weak magnetic order has been documented extensively by neutron diffraction [3-7] and to a lesser extent by magnetic x-ray scattering [8]. It has not been observed reliably in the standard thermal, magnetic and transport properties, even not by employing sensitive measuring techniques. Neutron-diffraction experiments [3] have demonstrated that SMAF and superconductivity ($T_c \sim 0.5$ K) coexist.

Substitution studies have demonstrated that UPt$_3$ is close to an antiferromagnetic instability [2, 9]. By replacing Pt by isoelectronic Pd pronounced phase transition anomalies appear in the thermal and transport properties. Notably, the $\lambda$-like anomaly in the specific heat and the chromium-type anomaly in the electrical resistivity give evidence for an antiferromagnetic phase transition of the spin-density-wave type. At optimal doping (5 at.% Pd) $T_{N,max} = 5.8$ K and the ordered moment equals $0.6 \pm 0.2$ $\mu_B$/U-atom [7, 10]. In order to distinguish this phase from the SMAF of pure UPt$_3$ we have termed it the large-moment antiferromagnetic (LMAF) phase. The magnetic instability can also be triggered by substituting Th for U [11-13]. Remarkably, the magnetic phase diagrams for the (U,Th)Pt$_3$ and U(Pt,Pd)$_3$ pseudobinaries are almost identical. This shows that the localisation of the uranium moments is not governed by the unit cell volume of these pseudobinaries (the unit cell volume decreases by Pd doping, while it increases by alloying with Th). Long-range magnetic order also shows up when UPt$_3$ is doped with 5 at.% Au, while substituting 5 at.% Ir, Rh, Y, Ce or Os, does not induce magnetic order [14-16]. This indicates that a shape effect, i.e. the change in the c/a ratio, is the relevant control parameter for the occurrence of magnetic order.

The magnetic phase diagram of the U(Pt$_{1-x}$Pd$_x$)$_3$ series has been measured [7] by neutron-diffraction and is shown in Fig. 1. The neutron-diffraction experiments have been performed on single-crystalline samples for $x \leq 0.05$ and the principal results are summarised as follows. The SMAF reported for pure UPt$_3$ is robust upon alloying and persists till at least $x= 0.005$. The ordered moment grows from $0.018 \pm 0.002$ $\mu_B$/U-atom for $x= 0$ to $0.048 \pm 0.008$ $\mu_B$/U-atom for $x= 0.005$. $T_N \approx 6$ K and, most remarkably, does not vary with Pd contents. Near $x= 0.01$ LMAF emerges. The ordered moment of this phase grows rapidly with Pd content and attains a maximum value of $0.63 \pm 0.05$ $\mu_B$/U-atom for $x= 0.05$ where also $T_N= 5.8$ K is maximum. $T_N(x)$ for the LMAF follows a Doniach-type phase diagram [17]. From this diagram it has been inferred that the antiferromagnetic instability for the LMAF in U(Pt$_{1-x}$Pd$_x$)$_3$ is located in the range 0.5-1 at.% Pd. It is important to notice that the neutron-diffraction data are consistent with a single-q magnetic structure which is identical for both the SMAF and LMAF phases. However a triple-q structure for the SMAF phase can not be excluded [6,7].

In this paper we report a μSR study of the evolution of magnetism in U(Pt,Pd)$_3$. This work was carried out in parallel with the neutron-diffraction study [7]. Our main objectives were: (i) to investigate the evolution of the weak magnetic order as function of Pd content, and (ii) to investigate the connection (or possible coexistence) between SMAF and LMAF. The motivation of using the μSR technique stems from the extreme sensitivity to magnetic signals. Besides the muon acts as a local probe, which permits to discern magnetically inequivalent sample regions. For recent reviews of μSR experiments on heavy-electron systems and magnetic materials we refer to Refs. 18 and 19.

Our work was in part inspired by the early μSR experiments on polycrystalline (U$_{1-x}$Th$_x$)Pt$_3$ reported by Heffner et al. [20]. For undoped UPt$_3$ these authors observed small increases of the Kubo-Toyabe relaxation rate, $\Delta_{KT}$, and the transverse-field Gaussian relaxation rate, $\sigma_G$, below $\approx 6$ K. These increases were attributed to weak static magnetism with a magnetic moment of the order of $10^{-3}$-$10^{-2}$ $\mu_B$/U-atom. This discovery in fact preceded the detection of SMAF by neutron diffraction. For U$_{0.95}$Th$_{0.05}$Pt$_3$, which orders antiferromagnetically at $T_N= 6.2$ K with a large magnetic moment ($\approx 0.6$ $\mu_B$/U-atom), spontaneous μ$^+$ oscillations at frequencies of 2 and 8 MHz were detected below $T_N$, while for U$_{0.99}$Th$_{0.01}$Pt$_3$ no spontaneous oscillations were observed ($T> 4.2$ K).

In our μSR study we concentrate predominantly on the low-temperature magnetic properties of the doped compounds. One of the principal results is that we, quite unexpectedly, could not resolve the SMAF in the zero-field experiments carried out on polycrystalline U(Pt$_{1-x}$Pd$_x$)$_3$ with $x=$ 0, 0.002 and 0.005. Also Dalmas de Réotier et al. failed to detect the SMAF by μSR in their high-quality single-crystalline samples [21]. These results are at variance with the data reported in Ref. 20. For higher Pd concentrations ($x=$ 0.01, 0.02 and 0.05) we observe spontaneous μ$^+$ precession frequencies, similar to those reported for Th doped UPt$_3$ [20].

This paper is organised as follows. Section 2 is devoted to the experimental details, like the sample preparation process, the characterisation of the samples and some relevant parameters of the experimental set-up. In sections 3 and 4 we present the results of the zero-field (ZF) and low transverse field (TF= 0.01 T) measurements on the SMAF and LMAF states, respectively. In section 5 we present TF(= 0.6 T) experiments on single-crystalline U(Pt$_{0.95}$Pd$_{0.05}$)$_3$. In section 6 we discuss our results, while the summary is presented in section 7. Parts of these results were presented in Refs. 22 and 23.

## 2. Experimental

The U(Pt$_{1-x}$Pd$_x$)$_3$ pseudobinaries crystallise in a hexagonal closed-packed structure (MgCd$_3$-type) with space group P6$_3$/mmc. The lattice parameters for $x=$ 0 are given by a= 5.764 Å and c= 4.899 Å. The lattice parameter for the a-axis for $x\leq$ 0.05 is constant within the experimental accuracy, while the lattice parameter for the c-axis decreases at a rate of 3x10$^{-4}$ Å per at.% Pd. This results in a minute reduction of the c/a ratio.

We have prepared polycrystalline material with $x=$ 0.000, 0.002, 0.005, 0.01, 0.02 and 0.05 by arc-melting the constituents in a stoichiometric ratio in an arc furnace on a water-cooled copper crucible under a continuously Ti-gettered argon atmosphere (0.5 bar). Samples with low Pd contents ($x\leq$ 0.01) were prepared by using appropriate master alloys (e.g. 5 at.% Pd). As starting materials we used natural uranium (JRC-EC, Geel) with a purity of 99.98%, and platinum and palladium (Johnson Matthey) with a purity of 99.999%. For annealing, the samples were wrapped in tantalum foil and put in water free quartz tubes together with a piece of uranium that served as a getter. After evacuating ($p<$ 10$^{-6}$ mbar) and sealing the tubes, the samples were annealed at 950 ºC during seven days. Next the samples were slowly cooled in three days to room temperature. For $x=$ 0.05 a single crystalline sample was pulled from the melt using a modified Czochralski technique in a tri-arc furnace under a continuously Ti-gettered argon atmosphere. The single-crystalline sample was annealed in a similar way as the polycrystalline material.

Four thin slices (thickness 0.8 mm, area 6x10 mm$^2$) were cut from the annealed polycrystalline buttons ($x$= 0, 0.002, 0.005, 0.01, 0.02 and 0.05) by means of spark-erosion. The surface layer, defected by spark-erosion, was removed by polishing with diamond paste (grain size 0.3 µm). The samples were glued on a silver support by General Electric varnish as to cover the desired area for the µSR experiments: 12x20 mm$^2$. The single-crystalline sample ($x$= 0.05) was glued to a silver rod, which served as sample support.

Parts of the polycrystalline batches were characterised by electrical resistivity measurements. In agreement with the data presented in Ref. 24, the upper superconducting transition temperature $T_c^+$ amounts to 0.533 and 0.389 K, for $x$= 0 and 0.002, respectively. Also the residual resistivity, $\rho_o$, was found to increase linearly with Pd contents (x≤ 0.005), which indicates that Pd dissolves homogeneously in the matrix. We obtain $\rho_o$ values of 0.88, 2.49, 6.2 and 12.0 µΩcm for $x$= 0.000, 0.002, 0.005 and 0.01, respectively. For higher Pd concentrations $\rho_o$ rises more rapidly because of the spin-density-wave type of magnetic order. The Néel temperatures of the 2 and 5 at.% Pd polycrystalline sample determined by resistivity amount to 4.0 and 6.3 K, respectively. These values are slightly higher than measured previously on other batches (3.6 and 5.8 K) [2].

The µSR experiments were performed at the Paul Scherrer Institute (Villigen), using the µ$^+$SR-dedicated beam-line πM3. ZF and TF data were collected at the General Purpose Spectrometer (GPS) using a $^4$He flow cryostat ($T$> 1.6 K). Here also the angular variation of the Knight shift was measured using an automated stepping motor device. Additional ZF and TF data were taken at the Low Temperature Facility (LTF), which is equipped with a top-loading $^3$He-$^4$He dilution refrigerator (Oxford Instruments) with a base temperature of 0.025 K. By changing the operation mode of the dilution refrigerator temperatures up to ≈10 K can be reached.

## 3. µSR experiments on SMAF compounds ($x$≤ 0.005)

Zero-field µSR experiments have been performed on polycrystalline U(Pt$_{1-x}$Pd$_x$)$_3$ samples with $x$= 0.000 in the temperature ($T$) interval 2.7-7.0 K, with $x$= 0.002 in the $T$-interval 0.9-8.0 K and with $x$= 0.005 in the $T$-interval 0.03-10 K. Some typical muon depolarisation curves, taken on the x= 0.005 compound at $T$= 0.1 K and 9.0 K, are shown in Fig. 2. For $x$≤ 0.005, the muon depolarisation for $T$< 10 K is best described by the standard Kubo-Toyabe function:

$$G_{KT}(\Delta_{KT}t) = \frac{1}{3} + \frac{2}{3}(1 - \Delta_{KT}^2 t^2)\exp(-\frac{1}{2}\Delta_{KT}^2 t^2) \qquad (1)$$

Here $\Delta_{KT} = \gamma_\mu \sqrt{<B^2>}$ is the Kubo-Toyabe relaxation rate, with $\gamma_\mu$ the muon gyromagnetic ratio ($\gamma_\mu/2\pi$= 135.5 MHz/T) and $<B^2>$ the second moment of the field distribution at the muon site. The Kubo-Toyabe function describes the case of an isotropic Gaussian distribution of static internal fields centred at zero field. The solid line in Fig. 2 presents a fit to the Kubo-Toyabe function for $x$= 0.005. In Figs. 3, 4 and 5 $\Delta_{KT}(T)$ is plotted for $x$= 0.000, 0.002 and 0.005, respectively. We conclude that $\Delta_{KT}$ shows no significant temperature dependence. The average values of $\Delta_{KT}$ are 0.065±0.005, 0.058±0.009 and 0.083±0.004 $\mu s^{-1}$ for $x$= 0.000, 0.002 and 0.005, respectively. Additional data for $x$= 0.002 were taken in a transverse field (perpendicular to the muon spin direction) of 0.010 T. Best fits were obtained using a Gaussian damped oscillation: $G(t)= \cos(2\pi\nu t+\phi)\exp(-½(\sigma_G t)^2)$. The Gaussian linewidth, $\sigma_G$, equals 0.081±0.007 $\mu s^{-1}$ and is temperature independent as well (see Fig. 4).

Surprisingly, our data for polycrystalline UPt$_3$ are at variance with the results reported by Heffner et al. [20] who observed a doubling of $\Delta_{KT}$, from 0.06 $\mu s^{-1}$ just above 6 K to 0.12 $\mu s^{-1}$ in the limit $T\rightarrow$ 0 K. As the doubling of $\Delta_{KT}$ was attributed to the presence of weak magnetic order (according to neutron diffraction SMAF), we conclude that the weak magnetic order does not show up in the ZF μSR signals for $x\leq$ 0.005. Is does also not show up in the TF= 0.010 T data. At this point it is important to realise that the neutron-diffraction experiments [7] show that for $x\leq$ 0.005 SMAF invariably sets in at $T_N\approx$ 6 K, while the ordered moment grows with Pd content: $m$= 0.018±0.002 $\mu_B$/U-atom, 0.024±0.003 $\mu_B$/U-atom and 0.048±0.008 $\mu_B$/U-atom for $x$= 0.000, 0.002 and 0.005, respectively. One could argue that the occurrence of SMAF is related to the single-crystalline nature of the samples used for neutron diffraction. However, the single and polycrystalline samples were prepared using the same high-purity starting materials and also the $\rho_0$-values are about the same. Our unexpected result is in agreement with recent experiments on high-purity single-crystalline UPt$_3$ carried out by Dalmas de Réotier et al. [19, 21]. These authors did not detect SMAF in their ZF μSR data, while neutron-diffraction measurements carried out on parts of the samples show that SMAF is present with the usual characteristics. Two explanations for the absence of SMAF in the μSR signals are conceivable: (i) the muons stop at sites where the dipolar fields due to the SMAF and the direct contact field cancel, and (ii) the antiferromagnetic moment fluctuates at a rate >10 MHz, i.e. too fast to be detected by μSR, but slower than the time scale of the neutron-diffraction experiment $\approx$ 0.1 THz. We come back to this most important issue in section 6.

The zero-field data for $x \leq 0.005$ (Figs. 3-5) can be attributed entirely to the depolarisation of the muon due to static $^{195}$Pt nuclear moments. For pure UPt$_3$ and U(Pt$_{0.998}$Pd$_{0.002}$)$_3$ experiments in a small (0.010 T) longitudinal field (along the muon spin direction) confirmed the static origin. We have calculated $\Delta^2_{KT}$ due to nuclear moments using the expression [25, 26]:

$$\Delta^2_{KT} = \frac{1}{6}\sum_j I_{Pt}(I_{Pt}+1)\left(\frac{\mu_0}{4\pi}\gamma_\mu \gamma_{Pt} \hbar\right)^2 \frac{5-3\cos^2(\theta_j)}{r_j^6} \quad (2)$$

Here the sum is over all $^{195}$Pt nuclei (abundance 33.7%) with spin $I_{Pt}=1/2$ and gyromagnetic ratio $\gamma_{Pt}$ ($\gamma_{Pt}/2\pi = 8.781$ MHz/T), which are located at a distance $r_j$ from the muon localisation site at an angle $\theta_j$ with respect to the muon spin ($\mu_0$ is the permeability of free space). For the most probable muon localisation sites the calculated values of $\Delta_{KT}$ range between 0.05 and 0.08 $\mu s^{-1}$ (see Table I). These calculations were performed for pure UPt$_3$, but for small amounts of Pt substituted by Pd, which has no nuclear moment, the corrections can be neglected. Since the measured values of $\Delta_{KT}$ also fall in the range 0.05-0.08 $\mu s^{-1}$ one cannot determine the stopping site from the depolarisation due to the nuclear moments.

## 4. LMAF probed by µSR experiments for $x \geq 0.01$

Zero-field µSR experiments have been performed on polycrystalline U(Pt$_{1-x}$Pd$_x$)$_3$ samples with $x=0.01$ in the $T$-interval 0.03-3 K, with $x=0.02$ in the $T$-interval 1.6-8.0 K and with $x=0.05$ in the $T$-interval 3.0-10 K. Additional experiments on a single-crystalline x=0.05 sample confirm the results obtained on the polycrystal. For all samples we can identify a magnetic phase transition temperature, where below a spontaneous $\mu^+$ precession frequency appears. This phase transition, which takes place at 1.8, 4.1 and 6.2 K for $x=0.01$, 0.02 and 0.05, respectively, is to the LMAF state. This is confirmed by the neutron-diffraction study on single-crystalline U(Pt$_{1-x}$Pd$_x$)$_3$, from which it follows that the Néel temperature equals 1.8, 3.5 and 6.2 K, and the ordered moment equals 0.11±0.03, 0.35±0.05 and 0.63±0.05 $\mu_B$/U-atom, for $x=0.01$, 0.02 and 0.05, respectively [7].

Since the magnetic behaviour does not vary strongly for $0.01 \leq x \leq 0.05$, we have fitted the µSR spectra of the three LMAF compounds with one and the same expression. Previously we have fitted the data of the $x=0.05$ compound with two spontaneous frequency components [22], while the data for $x=0.01$ have been analysed using one spontaneous frequency component and one exponential damping term [23]. The analysis with two terms, namely a standard depolarisation function for a polycrystalline magnet, $G_v(t)$, and a Lorentzian Kubo-Toyabe function, $G_{KL}(\lambda_{KL}t)$, yields a more consistent description when all Pd concentrations ($x$= 0.01, 0.02, 0.05) are considered. Below $T_N$ good fits are obtained using the following depolarisation function:

$$G(t) = A_1 G_v(t) + A_2 G_{KL}(\lambda_{KL}t) \tag{3a}$$

where

$$G_v(t) = \frac{2}{3}\exp(-\lambda t)\cos(2\pi\nu t + \phi) + \frac{1}{3}\exp(-\lambda' t) \tag{3b}$$

$$G_{KL}(\lambda_{KL}t) = \frac{1}{3} + \frac{2}{3}(1-\lambda_{KL}t)\exp(-\lambda_{KL}t) \tag{3c}$$

The Lorentzian Kubo-Toyabe term (eq. 3c) represents an isotropic Lorentzian distribution of internal fields with an average zero field. At the moment we do not have a satisfactory explanation why the most consistent analysis requires the Lorentzian Kubo-Toyabe term. It implies that the spectral distribution of the internal fields is better approximated by a Lorentzian rather than a Gaussian field distribution. In this respect the use of eq. 3 may be considered as a phenomenological approach. Consequently, the deduced parameters are phenomenological. Eq. 3 assumes two magnetically different muon stopping sites. Note that two magnetic sites not necessarily require two crystallographically different sites.

In the paramagnetic state ($T > T_N$) the muon depolarisation is best described by the standard Kubo-Toyabe function $G_{KT}(\Delta_{KT}t)$ (eq. 1), just as for the SMAF compounds ($x \leq 0.005$), with values of $\Delta_{KT}$ comparable to the values reported in Figs. 3-5. For $T \ll T_N$ we find $A_1 = A_2$. This suggests that half of the muons stops at sites where the dipolar fields cancel, while the other half stops at sites with a net local dipolar magnetic field.

In order to show that eq. 3 accounts for the LMAF state we have plotted in Fig. 6 and Fig. 7 the spontaneous frequency $\nu(T)$ and the depolarisation rate of the Lorentzian Kubo-Toyabe function, $\lambda_{KL}(T)$, respectively. For the LMAF state we found that the order parameter $m(T)$ as measured by neutron diffraction [7] could be described by $f(T)=f(0)(1-(T/T_N)^\alpha)^\beta$. The same expression, with almost identical values of $\alpha$ and $\beta$, yields a proper description of $\nu(T)$ and $\lambda_{KL}(T)$ as well (see solid lines in Fig. 6 and Fig. 7). The fit parameters are listed in Table II. For $x$= 0.02 and 0.05 the values of $\beta$ are close to the theoretical value $\beta$= 0.38 for the 3D Heisenberg model [27]. The phenomenological parameter $\alpha$ reflects spin-wave excitations. In a cubic antiferromagnetic system $\alpha$ is predicted to be 2 [28]. To our knowledge no predictions are available for a hexagonal system. A point of concern is that for a simple polycrystalline magnet one expects the spontaneous frequency $\nu(0)$ to scale with the ordered moment, which is clearly not the case here, as follows from the data in Table II. However, $\lambda_{KL}(0)$ scales with the ordered moment. We comment on this point in the next paragraph.

In order to demonstrate the relative weight of the terms in eq. 3 to the total depolarisation function, we have plotted $G_\nu(t)$ and $G_{KL}(\lambda_{KL}t)$ in Fig. 8 for a typical μSR spectrum in the LMAF state, taken on U(Pt$_{0.99}$Pd$_{0.01}$)$_3$ at $T=0.1$ K. In Fig. 9 the concentration dependence of $G_\nu(t)$ and $G_{KL}(\lambda_{KL}t)$ is shown. Whereas $G_{KL}(\lambda_{KL}t)$ varies smoothly with Pd contents, $G_\nu(t)$ is almost identical for $x=0.02$ and 0.05. If both signals were to originate from the same ordered moment, then $\lambda_{KT}$ and $\nu$ should both increase proportionally to the ordered moment. A possible reason for the absence of scaling of the frequency with the ordered moment is that the expression for $G_\nu(t)$ is only valid when $\lambda \ll \omega = 2\pi\nu$. Such a situation is described in Ref. 29 for a Gaussian field distribution. When $\lambda$ is of the same order as $\nu$ large systematic errors can influence the fit parameters of $G_\nu(t)$. Since we deal with heavily damped spontaneous oscillations this is in part the case. The temperature dependence of $\lambda$ is plotted in Fig. 10, which shows that $\lambda$ is almost constant below $T_N$ for $x=0.02$ and 0.05, with average values of 5.2±0.5 μs$^{-1}$ and 6.1±0.5 μs$^{-1}$, respectively. The fit to eq. 3a results in values $\lambda/\omega \approx 0.1$. Using the procedure as described in Ref. 29 we arrive at a correction of $\nu$ of only ≈ 10%. This small correction cannot explain the absence of scaling between $\nu$ and the ordered moment.

## 5. Transverse field μSR on U(Pt$_{0.95}$Pd$_{0.05}$)$_3$

In an attempt to determine the muon localisation site in the U(Pt$_{1-x}$Pd$_x$)$_3$ pseudo-binaries we have carried out μSR experiments for $x=0.05$ in transverse fields $B_{ext}=0.6$ T in the temperature interval 10-250 K. The sample was shaped into a cube (dimensions 5x5x5 mm$^3$) with edges along the principal crystallographic directions. Let us first focus on the low-temperature data. In order to determine the frequency components in the spectra for $\mathbf{B}_{ext}\parallel \mathbf{a}$ and $\mathbf{B}_{ext}\parallel \mathbf{c}$ at 10 K we have calculated the Fourier transforms, shown in Fig. 11. Inspecting the Fourier transforms we notice a remarkable difference for $\mathbf{B}_{ext}\parallel \mathbf{a}$ and $\mathbf{B}_{ext}\parallel \mathbf{c}$. Besides the background signal at 81.55 MHz, which is due to muons stopping in the silver sample support, we observe two signals for $\mathbf{B}_{ext}\parallel \mathbf{a}$ but only one signal for $\mathbf{B}_{ext}\parallel \mathbf{c}$. For $T>10$ K, the two signals for $\mathbf{B}_{ext}\parallel \mathbf{a}$ are no longer observed in the Fourier transform. Instead, a single but always asymmetric peak is observed.

Therefore we have analysed the spectra for $\mathbf{B}_{ext}\parallel \mathbf{a}$ ($T=10$-250 K) with the following three-component depolarisation function, $G_a(t)$:

$$G_a(t) = A_1 e^{-\lambda_1 t}\cos(2\pi\nu_1 t + \varphi) + A_2 e^{-\lambda_2 t}\cos(2\pi\nu_2 t + \varphi)$$
$$+ A_{bg} e^{-\lambda_{bg} t}\cos(2\pi\nu_{bg} t + \varphi) \quad (4)$$

The first two components account for the µSR signal from the sample and the third component is the background signal. Although the first two signals are not resolved in the frequency domain for $T > 10$ K, it is possible to fit both components in the time domain by fixing $A_1 = A_2$. In eq. 4 the envelops of the oscillating functions are exponentials. We also tried Gaussian damping, but it was not possible to discriminate between exponential or Gaussian damping terms. The resulting frequencies $\nu_1$ and $\nu_2$ are almost not influenced by the choice of the envelop function. For $\mathbf{B}_{ext}\|\mathbf{c}$ we have analysed the spectra using the following two-component depolarisation function, $G_c(t)$:

$$G_c(t) = A e^{-\lambda t}\cos(2\pi\nu t + \varphi) + A_{bg} e^{-\lambda_{bg} t}\cos(2\pi\nu_{bg} t + \varphi) \quad (5)$$

From the measured frequencies we have determined the Knight shift $K = (B_{loc}/B_{ext}) - 1 = (\nu/\nu_0) - 1$, where $B_{loc}$ is the field at the muon localisation site and $\nu_0 = \gamma_\mu B_{ext}/2\pi$. The Knight shift components for $\mathbf{B}_{ext}\|\mathbf{a}$, $K_{a,\nu 1}(T)$ and $K_{a,\nu 2}(T)$, and for $\mathbf{B}_{ext}\|\mathbf{c}$, $K_{c,\nu}(T)$, follow a Curie-Weiss behaviour in a limited $T$-range only (50-100 K). For $T > 100$ K the difference, $\nu_1 - \nu_2$, between the two frequencies ($\mathbf{B}_{ext}\|\mathbf{a}$) becomes smaller and above 150 K the data can only be fitted with one frequency. This indicates that slow muon hopping takes place for $T > 100$ K. As the muon diffuses, it experiences an average local magnetic field.

Next we compare the Knight shift with the susceptibility in order to determine the dipolar tensor components, $A^{ii}$. The susceptibility was measured on a different crystal using a SQUID magnetometer. For $T > 30$ K $\chi(T)$ follows a modified Curie-Weiss law, $\chi = \chi_0 + C/(T-\theta)$. The data are in excellent agreement with the results reported in Ref. 30. In Fig. 12 we present $K_i$ as function of the bulk susceptibility, $\chi_i$, in the so called Clogston-Jaccarino plot, $K_i(\chi_i)$, where the temperature is an implicit parameter. Here $\mathbf{B}\|\mathbf{i}$, where $\mathbf{i}$ denotes the principal crystallographic directions. In general $K_i$ scales with the susceptibility, $K_i = K_{con} + A^{ii}\chi_i$. $K_{con}$ is the Knight shift due to the direct contact field induced by the polarisation by the conduction electrons. The Clogston-Jaccarino plot shows several remarkable features: (i) $K_a(\chi_a)$ deviates strongly from the expected linear behaviour, while $K_c(\chi_c)$ is approximately linear, and (ii) the direct contact contribution to the Knight shift ($T \to \infty$) $K_{con}$ seems to be strongly anisotropic, while the unrenormalised Pauli susceptibility $\chi_0$ is not. This strongly suggests that the local and bulk susceptibilities differ, which hampers the determination of the components of the dipolar tensor.

The angular variation of the Knight shift $K(\theta)$ measured on a spherical U(Pt$_{0.95}$Pd$_{0.05}$)$_3$ sample [31] in the a-c plane follows a standard cos$^2$-law, while $K(\theta)$ in the basal plane is isotropic. From this it follows that the muon localisation site is restricted to axial symmetry. By analysing the $K(T)$-data in the temperature range 50-100 K with a modified Curie-Weiss type of expression, it has been concluded that there is only one stopping site, namely the 2a site, (0,0,0) [31]. The analysis of the Knight shift data taken on pure UPt$_3$, also points to the 2a site as stopping site [32]. This shows that the muon localisation site does not change at these small Pd concentrations.

The presence of two frequency components (**B**∥ **a**) and only one stopping site, provides strong evidence for two spatially distinct regions of different magnetic response up to at least ~100 K. This is in-line with a similar observation recently made for pure UPt$_3$ [33]. Whether the different magnetic response originates from macroscopically separated regions (e.g. domains) or is periodic in nature (e.g. a structural modulation [34]) remains an open problem.

## 6. Discussion

One of the unexpected conclusions from the present work is that SMAF ($x \leq 0.005$) is not detected in the zero-field µSR experiments. A first natural explanation of this result is that the SMAF is not present at all. However, this is contradicted by neutron-diffraction experiments. A simple calculation shows that the contribution to the dipolar field from the single-q (or possibly triple-q) magnetic structure cancels at the muon localisation site (0,0,0). This offers a second explanation for not detecting the SMAF. Indeed, the measured values of $\Delta_{KT}$ (see section 3) are consistent with depolarisation of the muon due to Pt nuclear moments only. Moreover, a comparison of the measured and calculated values of $\Delta_{KT}$ for axial symmetric stopping sites (Table I) is not inconsistent with the (0,0,0) stopping site. A third explanation for not observing the SMAF is that the antiferromagnetic moment fluctuates at a rate >10 MHz, i.e. too fast to be detected by µSR, but slower than the time scale of the neutron-diffraction experiment ≈ 0.1 THz. This explanation is particularly appealing because it also clarifies the absence of a signature of the SMAF in NMR experiments [35, 36]. Kohori et al. [35] carried out $^{195}$Pt NMR on the U(Pt$_{1-x}$Pd$_x$)$_3$ system. For compounds with LMAF (e.g. $x$= 0.05) zero-field experiments showed that the transferred hyperfine field does not cancel at the Pt site. A simple calculation showed that the internal field originating from SMAF for $x$= 0 should have been observed as well, however, this turned out not to be the case. Since the symmetry argument for cancellation of the dipolar field does not hold for the Pt sites, the most probable explanation for the absence of the SMAF is the one of the fluctuating moment.

For $x \geq 0.01$ the LMAF is clearly observed in the zero-field μSR data. The muon depolarisation in the ordered state is described by two terms of equal amplitude: an exponentially damped spontaneous oscillation and a Lorentzian Kubo-Toyabe function. However, it is not understood why the spontaneous frequency ($\nu$) does not scale with the ordered moment, while the linewidth $\lambda_{KL}$ does. Although the analysis of the Knight shift, measured in a transverse field of 0.6 T, shows that there is only one axial symmetric stopping site (0,0,0), we find two distinct values of $K_{con}$ for **B**$\|$ **a** (see Fig. 12 and Ref. 31). The observation of two contributions in the ordered state, i.e. one term with large spontaneous frequencies (in the range 4.7-8.1 MHz) and a second term described by the Lorentzian Kubo-Toyabe function, is possibly related to the large differences in $K_{con}$. At the moment we cannot offer an explanation how $K_{con}$ and the spontaneous dipolar fields below $T_N$ may be connected to each other, but the absence of scaling of the spontaneous frequency with the ordered moment $m$ might be another indication for an unusual muon depolarisation mechanism. High-resolution transverse-field experiments are needed to clarify these issues.

The μSR and neutron-diffraction studies demonstrate that SMAF and LMAF in the U(Pt$_{1-x}$Pd$_x$)$_3$ pseudobinaries are not closely connected. The differences between SMAF and LMAF are: (i) $T_N(x)$ attains a constant value of $\approx$ 6 K for SMAF, while $T_N(x)$ of the LMAF compounds follows a Doniach-type phase diagram, (ii) the squared order parameter $m^2(T)$ for the SMAF, as measured by neutron diffraction, grows in an unusual quasi-linear fashion, while the order parameter for the LMAF is conventional and confirms a real phase transition, and (iii) SMAF is not observed in zero-field μSR experiments in contrast to LMAF. The latter point we attribute to the fluctuating nature of the small ordered moment, which is consistent with NMR data [35, 36]. This strongly suggests that the SMAF does not present a true phase transition, but rather is a crossover phenomenon. The μSR and neutron-diffraction studies both show that LMAF is present for $x \geq 0.01$, while it is no longer observed for $x = 0.005$. This implies that the antiferromagnetic instability for the LMAF is located in the concentration range $x = 0.005$-$0.01$. μSR experiments on samples with intermediate Pd concentrations are underway in order to determine the critical concentration for LMAF, $x_c$. Of particular interest here is to investigate whether this critical concentration coincides with the critical concentration for the suppression of superconductivity $x_c = 0.007$ [24]. This would provide strong evidence that LMAF and superconductivity compete.

## 7. Summary

μSR experiments have been carried out on a series of pseudobinary polycrystalline heavy-electron U(Pt$_{1-x}$Pd$_x$)$_3$ compounds ($x \leq 0.05$). For $x \leq 0.005$ SMAF is not observed in the zero-field signals, whereas neutron diffraction shows that SMAF is stable upon alloying and $T_N(x) \approx 6$ K. The μSR spectra for $x \leq 0.005$ are consistent with depolarisation of the muon due to nuclear moments only. The absence of SMAF signals in the ZF data is attributed to the fluctuating nature of the small ordered moment, consistent with NMR data [35, 36]. This strongly suggests that the SMAF does not present a true phase transition, but rather is a crossover phenomenon. For $0.01 \leq x \leq 0.05$ LMAF is clearly observed in the zero-field μSR data. The muon depolarisation in the ordered state is described by two terms of equal amplitude: an exponentially damped spontaneous oscillation and a Lorentzian Kubo-Toyabe function. The depolarisation rate of the Lorentzian Kubo-Toyabe function, $\lambda_{KL}(T)$, was found to scale with the ordered moment $m(T)$ as measured by neutron diffraction. The Knight shift measured at 0.6 T on single-crystalline U(Pt$_{0.95}$Pd$_{0.05}$)$_3$ in the paramagnetic state shows two signals for **B** $\perp$ **c**, but only one signal for **B** $\parallel$ **c**. In order to reconcile the observation of two signals for **B** $\perp$ **c** with the presence of only one stopping site (0,0,0) [31], one has to evoke two spatially distinct regions of different magnetic response, like recently reported for pure UPt$_3$ [33].

## Acknowledgments

We thank R. van Harrevelt for contributing to this work in an early stage and C. Baines for skillful operating the LTF. A. Yaouanc is gratefully acknowledged for stimulating discussions. This work was part of the research program of the Dutch Foundation for Fundamental Research of Matter ("Stichting" FOM).

Table I   The calculated Kubo-Toyabe linewidths, $\Delta_{KT}$, of UPt$_3$ in the polycrystalline limit for axial symmetric sites. The first column gives the multiplicity and the Wyckoff letter of the particular site. The numbers between parentheses denote the error in averaging over the possible $^{195}$Pt configurations around the muon using the Monte Carlo method.

| site | | $\Delta_{KT}$ (µs$^{-1}$) |
|---|---|---|
| 2a | 0 0 0 | 0.061(1) |
| 2b | 0 0 1/4 | 0.081(1) |
| 4e | 0 0 1/8 | 0.073(1) |
| 4f | 2/3 1/3 0 | 0.046(1) |
| 2d | 2/3 1/3 1/4 | 0.079(1) |

Table II   Fitting parameters for the LMAF state, determined from the zero-field temperature dependences of $m$, $\lambda_{KL}$, and $\nu$, described by the relation $f(T) = f(0)(1-(T/T_N)^\alpha)^\beta$. The subscript ND refers to parameters determined from neutron-diffraction experiments, while the subscripts KL and $\nu$ refers to the parameters determined from the µSR data (see eq. 3). The numbers between parentheses denote the error.

| $x$ | $m(0)$ (µ$_B$) | $T_{N,ND}$ (K) | $\alpha_{ND}$ | $\beta_{ND}$ |
|---|---|---|---|---|
| 0.01 | 0.11(3) | 1.6(2) | - | - |
| 0.02 | 0.34(5) | 3.5(2) | 1.9(2) | 0.50(5) |
| 0.05 | 0.63(5) | 5.8(1) | 1.8(1) | 0.32(3) |
| $x$ | $\lambda_{KL}(0)$ (µs$^{-1}$) | $T_{N,KL}$ (K) | $\alpha_{KL}$ | $\beta_{KL}$ |
| 0.01 | 0.76(5) | 1.58(8) | 1.9(4) | 0.85(30) |
| 0.02 | 4.1(3) | 4.16(6) | 1.9(2) | 0.36(5) |
| 0.05 | 9.3(9) | 6.35(12) | 2.0(5) | 0.36(6) |
| $x$ | $\nu(0)$ (MHz) | $T_{N,\nu}$ (K) | $\alpha_\nu$ | $\beta_\nu$ |
| 0.01 | 4.7(2) | 1.75(9) | 1.5(4) | 0.48(9) |
| 0.02 | 7.9(1) | 4.15(1) | 2.0(2) | 0.39(2) |
| 0.05 | 8.1(1) | 6.21(1) | 2.1(3) | 0.39(2) |

**Figure captions**

Fig. 1   The Néel temperature, $T_N$, versus Pd concentration for U(Pt$_{1-x}$Pd$_x$)$_3$ alloys as determined from neutron diffraction (○), specific heat (□) and μSR (■) experiments. SMAF and LMAF denote small-moment and large-moment antiferromagnetic order, respectively. In the lower left corner the upper superconducting transition temperature as determined by resistivity experiments is given. SC denotes the superconducting phase. This neutron diffraction and specific heat data is taken from Ref. 7.

Fig. 2   Typical zero-field spectra measured for polycrystalline U(Pt$_{0.995}$Pd$_{0.005}$)$_3$. The solid line represents a fit to the Kubo-Toyabe function. The muon depolarisation is the same above and below the antiferromagnetic transition ($T_N \approx$ 6 K).

Fig. 3   Zero-field Kubo-Toyabe line width, $\Delta_{KT}$, for polycrystalline UPt$_3$. The solid line indicates the average value.

Fig. 4   Zero-field Kubo-Toyabe line width (○) and transverse field (0.01 T) Gaussian line width (●) for polycrystalline U(Pt$_{0.998}$Pd$_{0.002}$)$_3$. The solid lines indicate the average values.

Fig. 5   Zero-field Kubo-Toyabe line width for polycrystalline U(Pt$_{0.995}$Pd$_{0.005}$)$_3$. The solid line indicates the average value.

Fig. 6   Temperature variation of the spontaneous frequency ν for polycrystalline U(Pt$_{1-x}$Pd$_x$)$_3$ with $x=$ 0.01, 0.02 and 0.05 (see eq. 3). The solid lines represent fits to the function $f(T)=f(0)(1-(T/T_N)^\alpha)^\beta$ (see text).

Fig. 7   Temperature variation of $\lambda_{KL}$ for polycrystalline U(Pt$_{1-x}$Pd$_x$)$_3$ with $x=$ 0.01, 0.02 and 0.05 (see eq. 3). The solid lines represent fits to the function $f(T)=f(0)(1-(T/T_N)^\alpha)^\beta$ (see text). $\lambda_{KL}(T)$ scales with the ordered moment.

Fig. 8   Typical zero-field spectrum measured for polycrystalline U(Pt$_{0.99}$Pd$_{0.01}$)$_3$ at $T=0.1$ K. The lines represent the different components of the fit (see eq. 3).

Fig. 9   Contributions to the muon depolarisation function (see eq. 3) in the ordered state for polycrystalline U(Pt$_{1-x}$Pd$_x$)$_3$ with $x=$ 0.01, 0.02 and 0.05 at $T=$ 0.25, 2 and 3 K, respectively. Upper frame: Lorentzian Kubo-Toyabe function. Lower frame: exponentially damped oscillating component.

Fig. 10   Temperature variation of the damping parameter $\lambda$ of the oscillatory term for polycrystalline U(Pt$_{1-x}$Pd$_x$)$_3$ with $x=$0.01, 0.02 and 0.05 (see eq. 3).

Fig. 11   Fourier transforms of the spectra measured at 10 K in a transverse field of 0.6 T for single-crystalline U(Pt$_{0.95}$Pd$_{0.05}$)$_3$: solid line **B**∥ **a**, dotted line **B**∥ **c**. The arrow indicates the background signal.

Fig. 12   Clogston-Jaccarino plot for U(Pt$_{0.95}$Pd$_{0.05}$)$_3$: (●) one component fit for **B**∥ **a**, (■, ◆) two component fit ($\nu_1$, $\nu_2$) for **B**∥ **a**, and (○) **B**∥ **c**. The temperature is an implicit parameter (10 K < T < 250 K) with the low temperatures at the right side of the plot. The splitting for **B**∥ **a** disappears above 100 K.

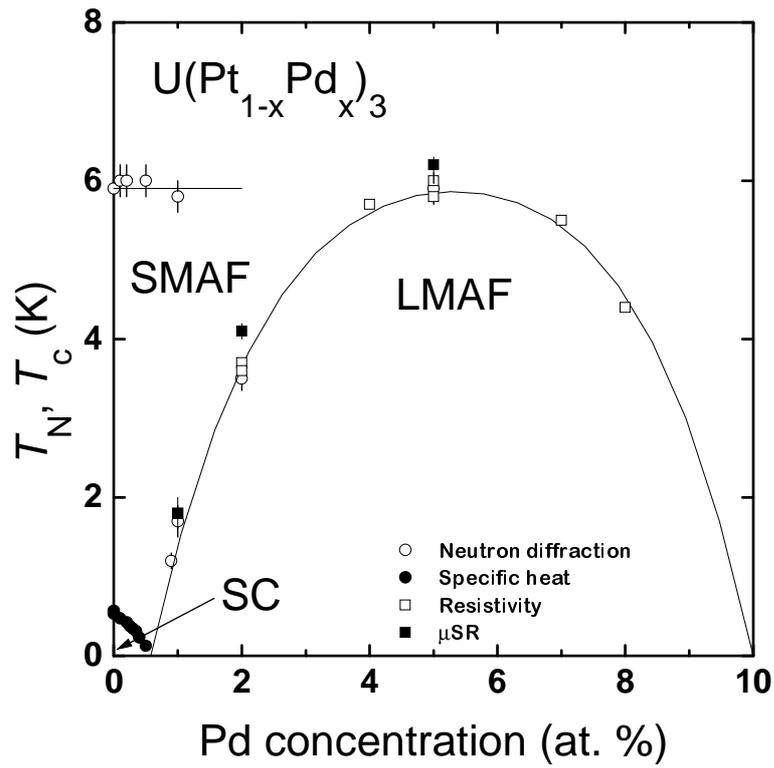

Fig. 1

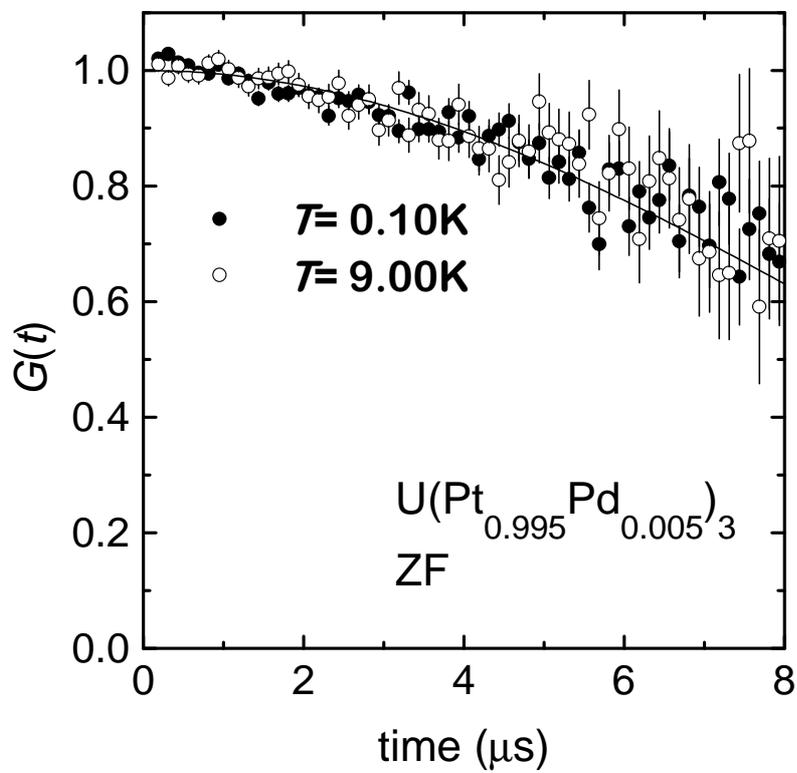

Fig. 2

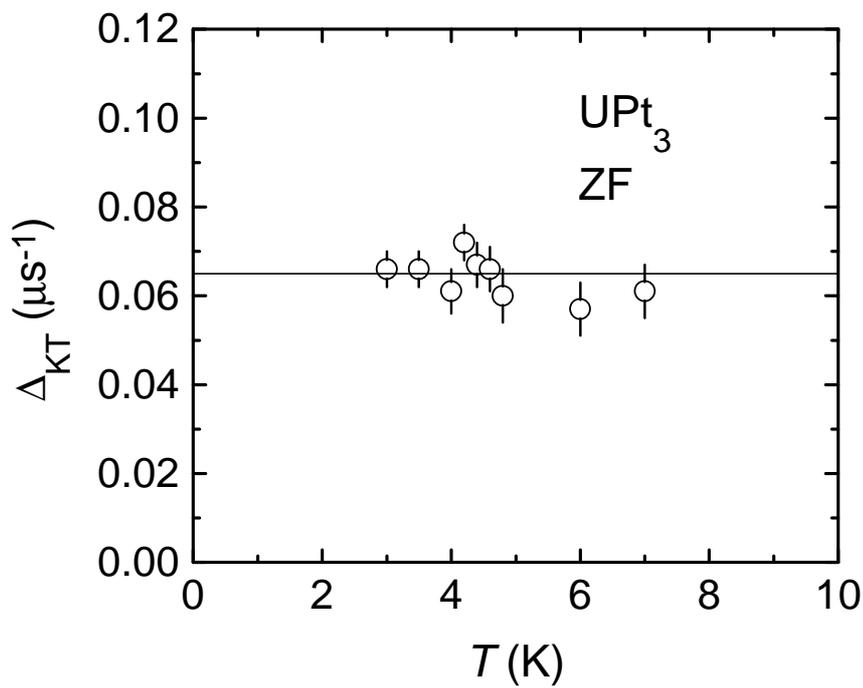

Fig. 3

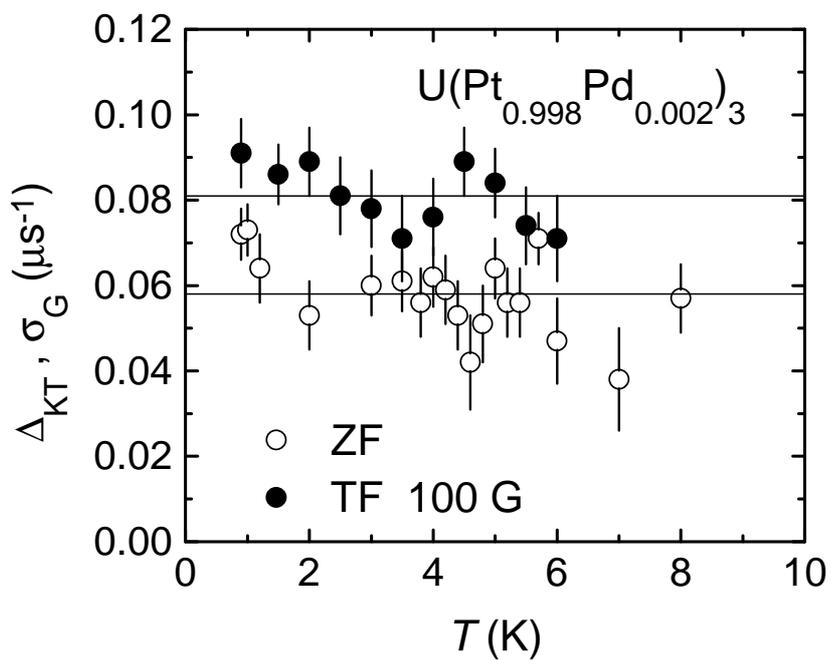

Fig. 4

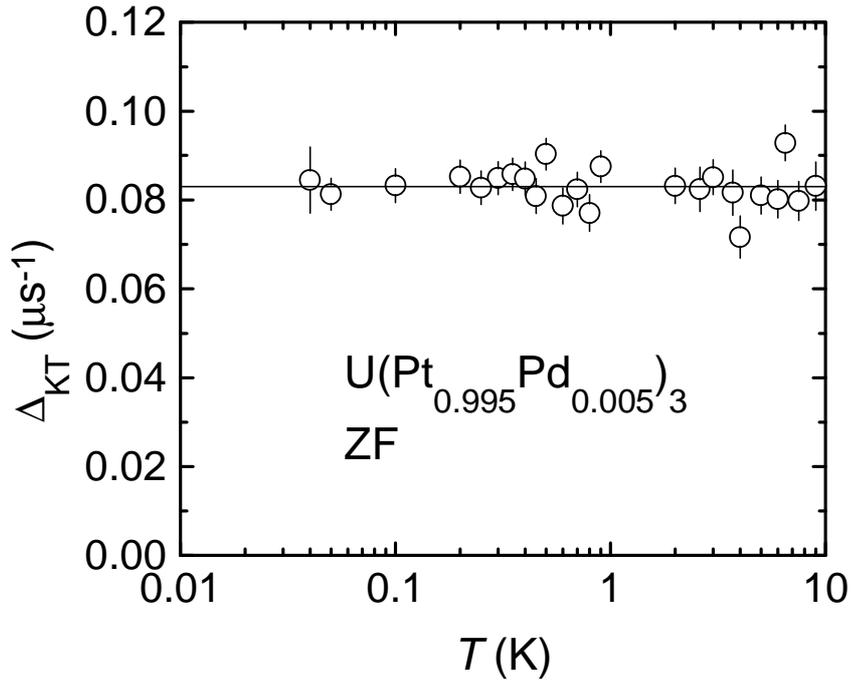

Fig. 5

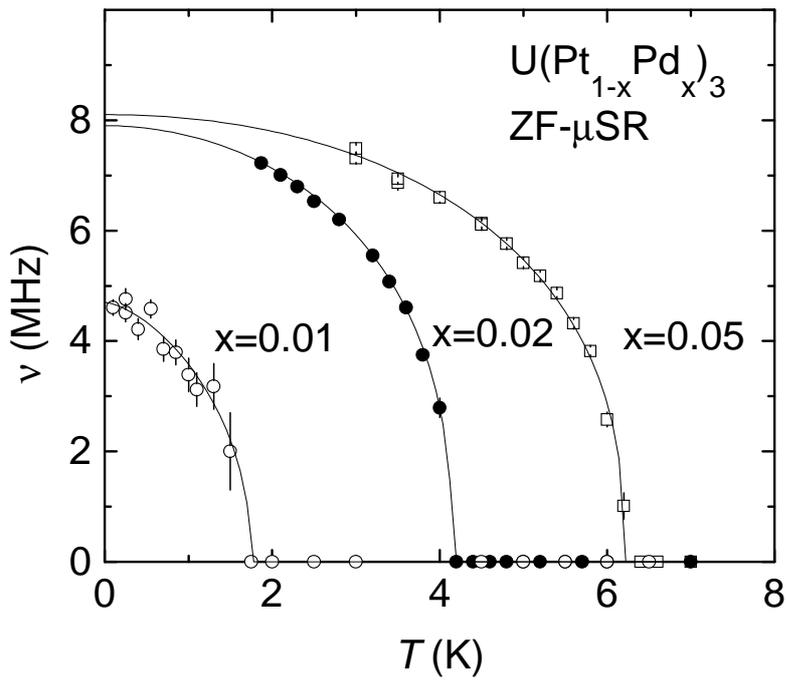

Fig. 6

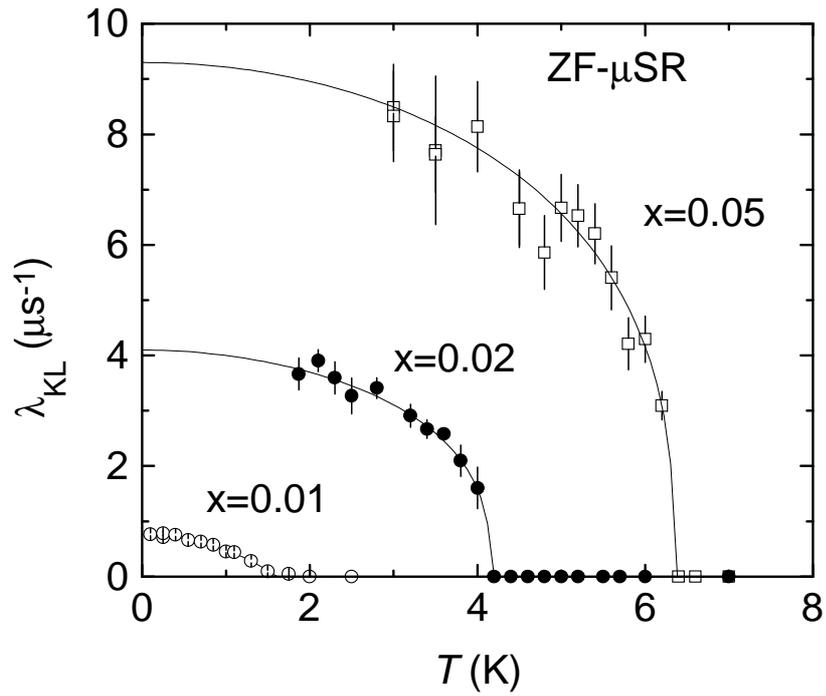

Fig. 7

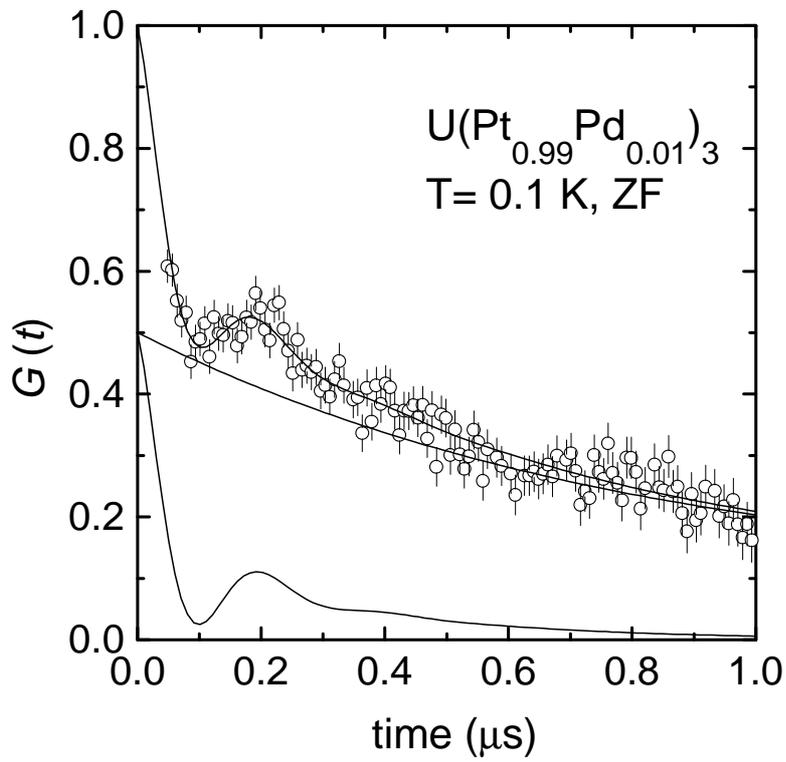

Fig. 8

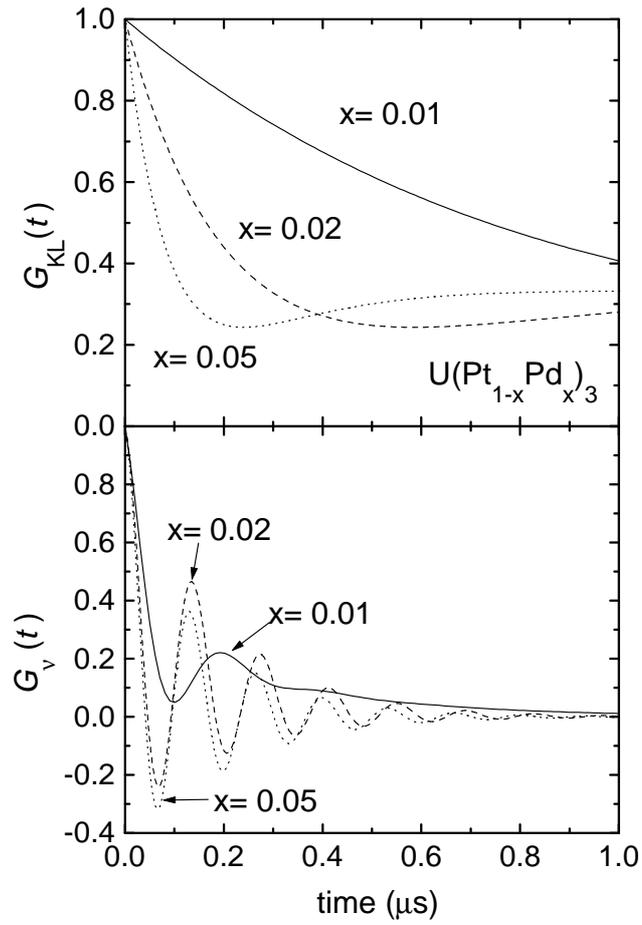

Fig. 9

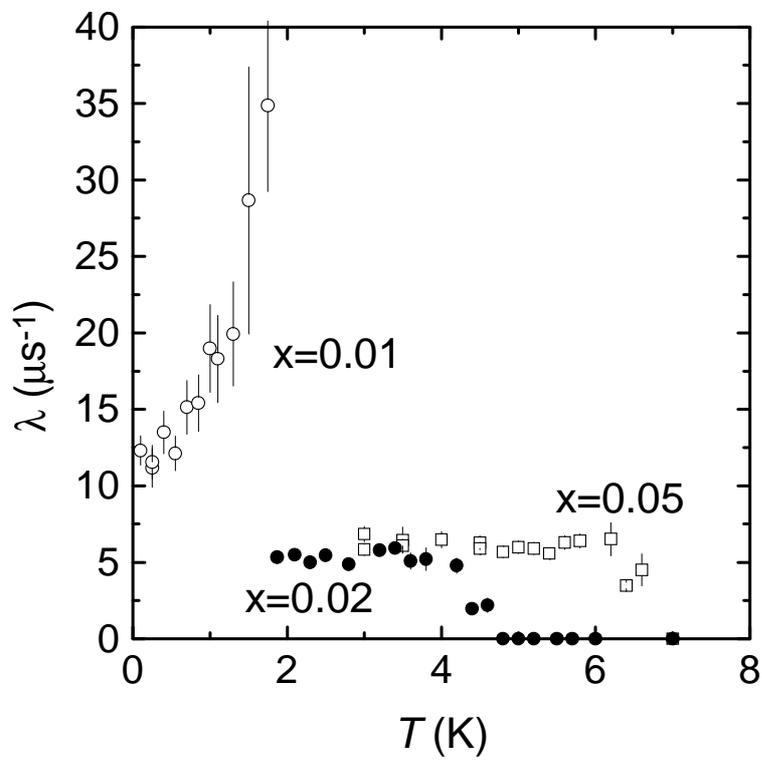

Fig. 10

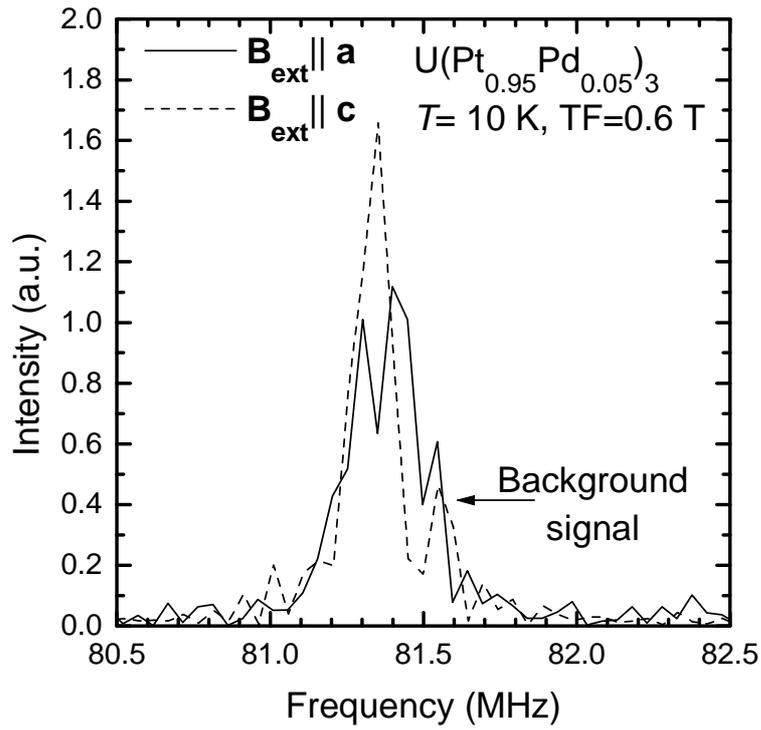

Fig. 11

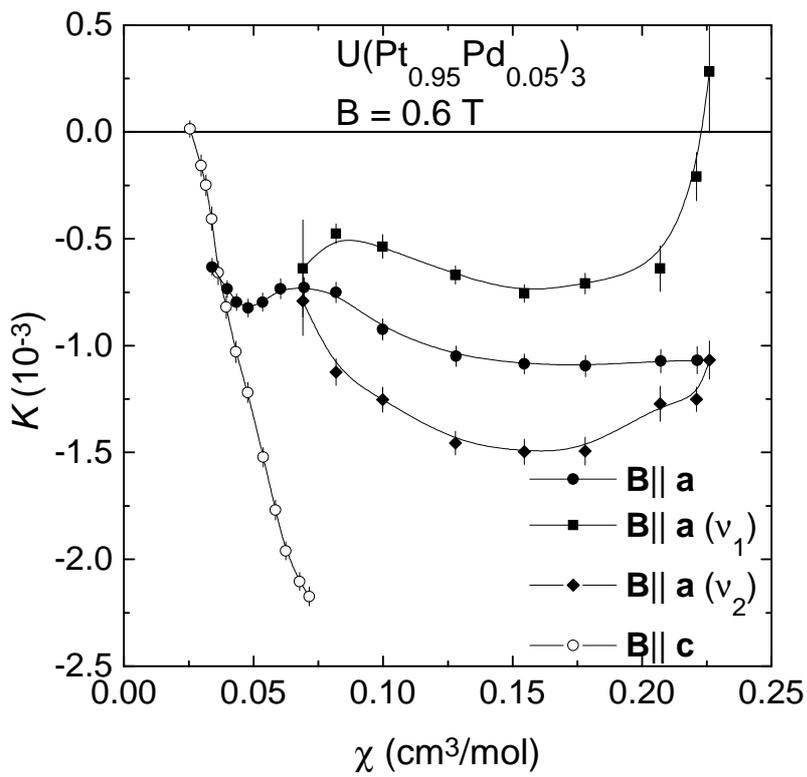

Fig. 12